\documentstyle[prl,aps,epsfig,12pt]{revtex}

\def\la{\langle}
\def\ra{\rangle}

\begin{document}
 \tolerance 50000

\draft

\title{Matrix Product Approach to Conjugated Polymers} 

\author{ 
M.A. Martin-Delgado$^{1}$, G. Sierra$^{2}$, S. Pleutin$^{3}$ and
E. Jeckelmann$^{4}$ 
}
\address{
$^{1}$Departamento de F{\'{\i}}sica 
Te{\'o}rica, Universidad Complutense, Madrid, Spain.\\
$^{2}$Instituto de Matem{\'a}ticas y F{\'\i}sica Fundamental, C.S.I.C.,
Madrid, Spain. \\
$^{3}$Max-Planck-Institut f\"{u}r Physik Komplexer 
Systeme, Dresden, Germany\\
$^{4}$Fachbereich Physik, Philipps-Universit\"{a}t Marburg, Marburg, Germany 
}

\maketitle 

\begin{abstract} 
\begin{center}
\parbox{14cm}{ The Matrix Product method (MPM) 
has been used  
in the past to generate variational ansatzs of the ground state (GS) 
of spin chains and  ladders.
In this paper  we apply the MPM  to study  the GS
of conjugated polymers in the valence bond basis, exploiting 
the charge and spin  conservation as well as the electron-hole
and spin-parity symmetries. We employ  the $U-V-\delta$ Hamiltonian
which is  a simplified version of the PPP Hamiltonian. 
For  several coupling constants $U$ and $V$  
and dimerizations $\delta$  we compute 
the GS energy per monomer 
which agrees within a $2\%-4\%$ accuracy with
the DMRG results.  We also show the evolution of the MP-variational
parameters in the weak and strong dimerization regimes. 
}

\end{center}
\end{abstract}

\pacs{
\hspace{2.5cm} 
 }

\section{Introduction}

The study of conjugated polymers has been a subject of great interest
for over two decades. There are both theoretical and technological reasons
for this interest \cite{review}.
On the theoretical side, there exists a controversy within the scientific 
community  over how to explain, understand and describe 
the photophysics/photochemistry of this class of materials. 
This 
controversy
 is of such a fundamental nature that the solution 
of the problem might be in a unification of the semiconductor
and metal physics with the molecular quantum chemistry. 
On the technological side, pi-conjugated polymers behave as semiconductors
and this has prompted several research groups
to investigate the physics of these materials in an effort to determine       
 their potential for improving the performance 
and efficiency and reducing the cost of light-emitting diodes (LEDs).
More recently, they are also considered to make an entrance in the 
field of photovoltaics, where they could be used as solar cells.

Saturated polymers
 are long chains of molecules, generally 
made of
carbon with
 hydrogen on the sides, all attached to one another by single bonds.
This constitutes the backbone of the macromolecule. The most relevant
feature of these structures is the fact that the bonds are all single bonds
or, in other words, that all the binding are of $\sigma$-- type.
Saturated polymers are then all very insulating; they are not 
electronically
interesting but are known for their flexibility although they are also quite
 mechanically strong materials. The most familiar of these compounds is the
polyethylene.

On the contrary, conjugated polymers show very interesting electronic
properties together with remarkable mechanical properties; for instance, they
can emit light and conduct electricity \cite{review}. In these compounds, two of the three
$2p$ orbitals on each carbon atom hybridize with the $2s$ orbital to form
three $sp^2$ molecular orbitals. These orbitals are responsible for the
backbone of the molecular chain; these are the so-called $\sigma$--orbitals. The third carbon orbital is $p_z$ and points perpendicular to
the chain. There exist a strong overlapping between nearest-neighbours $p_z$
orbitals so that the corresponding electrons are fully delocalized on the
whole molecule; these are the $\pi$ electrons responsible for all the
interesting electronic properties of low-energy. For instance, because of these electrons, the
linear chain, the polyacetylene, which is considered in this work, is
dimerized: its backbone shows an alternance between double and single
bonds. Quite 
generally
, despite a huge amount of works, the electronic properties of these compound stay rather
controversial \cite{review}.

The delocalization character of the electrons in the $\pi$ molecular
orbitals of the conjugate polymer chains led to the introduction of
model Hamiltonians to study and predict their electronic properties.
The initially simplest 
possible
 model is a tight-binding approximation
or H\"{u}ckel model \cite{huckel} to describe the motion of $\pi$ 
electrons in a free way. This is a very crude approximation which
has been improved in several fashions. One of them is the inclusion
of electron-phonon interactions \cite{ssh,holstein}.
However, this is not enough as the electronic properties of $\pi$-conjugated
polymers 
derive from a
 true many-body problem were electron-electron interactions
are equally important as the electron-phonon interactions. Then, the
PPP 
(Pariser-Parr-Pople)
 Hamiltonian \cite{ppp1,ppp2,review} is used to model these 
electron-electron effects in
a first approximation without taking into account phonon effects, to
make simpler a first analysis of the electronic properties.
In the PPP Hamiltonian, the alternating single-double bonds of the 
backbone polymer structure is realized by means of a dimerization 
term in the hopping kinetic energy. The most general form of the 
PPP Hamiltonian reads as follows:

\begin{eqnarray}
H_{PPP} &=& H_{\rm K} + H_{\rm I}, \\
H_{\rm K} &=& - \sum_{\la i,j \ra,\sigma} t ( 1 - \delta (-1)^i)
( c_{i,\sigma}^\dagger c_{j,\sigma} + {\rm h.c.} ) \\
H_{\rm I} &=& U \sum_{i} 
(c_{i,\uparrow}^\dagger c_{i,\uparrow} - \frac{n_{\rm el}}{2})
(c_{i,\downarrow}^\dagger c_{i,\downarrow} 
- \frac{n_{\rm el}}{2}) \nonumber \\
&+& \sum_{i,j} V(r_{i,j}) 
(\sum_\sigma c_{i,\sigma}^\dagger c_{i,\sigma} - n_{\rm el})
(\sum_\tau c_{j,\tau}^\dagger c_{j,\tau} - n_{\rm el})
\end{eqnarray}

\noindent where $H_K$ is the dimerized tight binding kinetic part
and $H_I$ represents the Coulomb interactions among the electrons.
Here the operators $c_{i,\sigma}$, $c_{i,\sigma}^\dagger$
are standard creation and 
annihilation
 operators for $\pi$ electrons
at carbon site $i$ with spin $\sigma$.
The parameter $t$ is the hopping overlapping integral between the 
nearest-
neighbour
 carbon atoms,  $\delta$ measures the dimerization 
of the chain, $n_{\rm el}$, is the average number of electrons per site,
$U$ is the on-site Coulomb repulsion between the electrons,
$r_{i,j}$ is the distance between sites $i$ and $j$ along the chain
and $V(r_{i,j})$ is the long range contribution of the Coulomb repulsion.

The PPP Hamiltonian has been the subject of extensive studies using
a great variety of techniques such as 
 Hartree-Fock, CI calculations,
small cluster exact diagonalization, Quantum Monte Carlo 
and so on
and so forth \cite{review}.
Only recently it has become possible to apply a new numerical technique,
the Density Matrix Renormalization Group (DMRG) \cite{white},
which allows us to obtain highly accurate results both for small, intermediate
and large polymer chains \cite{Ramashesa1,ramasesha,bb,yaron,fano}. 
These DMRG studies have helped to clarify 
the correct ordering of excited states in the low energy part of the 
spectrum which are relevant for the nonlinear spectroscopic experiments.

In this paper we will concentrate on the study of the PPP Hamiltonian and
leave the  effect of interaction with phonons for future studies.
The PPP Hamiltonian has been studied using an excitonic method
based on a local description of the polymer in terms of monomers \cite{pleutin}.
The relevant electronic configurations are built on a small number of pertinent
local excitations. This has provided a simple and microscopic physical approximate
picture of the model.
Recently we have extended this local configuration studies using the Recurrent 
Variational Approach (RVA) method  \cite{RVA-poly} in order to study larger
polymer chains in a systematic way while retaining the 
previous intuitive physical picture.
The RVA \cite{RVA} is a non-perturbative variational method  in which 
one retains a single state as the best candidate
for the ground state of the system. This reduction of degrees of freedom
is initially done in order to keep the method manageable analytically.
The aim of this analytical approach
is to try to understand the relevant physical degrees of
freedom so that we can figure out what  the underlying physics is in a
strongly correlated system.
This initial analytical goal has been also developed in order to later
acquire more numerical precision. To do this, the method becomes more
numerical  and somehow stands in between an analytical formulation of the
DMRG and a numerical one.

\noindent This effort of understanding the relevant electronic configurations
in conjugated polymers has been also carried out in exact small cluster
calculations using excitonic Valence Bond basis \cite{exciton-basis} for polymer chains
of length up to 10 sites, arranged into diatomic 
ethylene
 molecules.

A first comparison of RVA results  with DMRG gave us promising perspectives to improve these
variational calculations \cite{RVA-poly} by incorporating more local configurations 
and variational parameters.  In this paper we undertake this project by using a
Matrix Product ansatz for the ground state (GS) wave functions \cite{matrix-product}.
This ansatz is a variational approach based on first order Recursion Relations (RR's)
instead of second order RR's as in the RVA \cite{RSDM}.
With this RR's we construct the GS of the polymer chain in different 
symmetry
 sectors
based on the 16 local configurations of the diatomic 
ethylene
 molecule within the
PPP approximation. Thus, the chain is built up by adding one
 ethylene
 at each step of the
variational process.

This paper is organized as follows.  In Section 2 we introduce a Matrix Product ansatz
specially adapted for the PPP Hamiltonian. In Section 3 we set up the Recurrent Relations
to compute the GS energies in several sectors according to prescribed symmetries.
In Section 4 we present  variational and DMRG results and make a comparison 
obtaining a very good agreement between them.
Section 5 is devoted to prospects and conclusions.

\section{The Matrix Product Ansatz}

 The main idea of the MP method is to generate 
the ground state (GS) of a quasi-one dimensional  system 
in terms of a set
of states $ |\alpha \rangle_N$ generated by  the following 
recursion formula \cite{matrix-product,RSDM},

\begin{equation}
|\alpha \rangle_N = \sum_{m, \beta} \;
A_{\alpha, \beta}[m] \; |m \rangle_N \, 
|\beta \rangle_{N-1} 
\label{1}
\end{equation}

\noindent where $N$ denotes the number of lattice sites and
$|m \rangle_N $ is  a set of states located at the site $N$.
For 
conjugated
 polymers each lattice site in (\ref{1}) 
refers  to a monomer unit, and hence  $|m \rangle_N$
describes  the 16 possible states associated to a 
single monomer. In table 1 we show the basis
of local monomer states $|m \rangle_N$ 
used in our construction. We have adopted a
valence bond basis which is more convenient to our
purposes although it can be easily related
to the exciton-valence bond basis of 
references \cite{RVA-poly,exciton-basis}.

The states $|\alpha \rangle_N$ have to be regarded
as block states made  of 
intricate
 combinations of 
$N$ monomeric states whose structure depends of the 
MP amplitudes $A_{\alpha, \beta}[m]$, 
which in fact are the variational parameters of the method.
The latter parameters can be made to depend on the
step $N$ of the RR, but in the thermodynamic limit
one expect them  to reach a fixed point value. Below we shall
assume the thermodynamic limit, i.e. independence
of $A_{\alpha,\beta}[m]$ on $N$,  although computations can
be done for any finite value of $N$.  
The choice of the block states $|\alpha \rangle_N$ is mainly
dictated by physical considerations and they are characterized
by a set of quantum numbers as spin, charge, etc. In the case
of conjugated polymers we shall keep 6 block states which 
are to be thought as the GS's in the following sectors of the Hilbert
space: i) singlet state at half filling with symmetry  
$^1A^+_g$, ii) singlet state at half filling with symmetry
$^1B^-_u$, iii) a spin 1/2 doublet corresponding to making a hole
to the half filled GS and iv) a spin 1/2 doublet corresponding
to the addition of one electron to the half filled GS. 
The last two cases iii) and iv) describe localized charge transfer 
excitations between monomers, which play an important role
in the GS of the polymer. In table 2 we give the 6 blocks used
in the MP-ansatz. 

Altogether
 we have a total of
$6 \times 16 \times 6 = 576$ 
possible MP amplitudes, but further constraints greatly 
reduce this number. 
First of all and without loose of generality
one can impose that the block states $|\alpha \rangle_N$
are 
orthonormal
. This is guaranteed, for any value
of $N$,  by  the following normalization
conditions on the $A's$,

\begin{equation}
\sum_{m,\beta} A_{\alpha,\beta}[m] \; A_{\alpha',\beta}[m]
= \delta_{\alpha, \alpha'}
\label{2}
\end{equation}

Moreover,  the RR (\ref{1}) should  preserve the charge
and the spin of the states, reflected in the equations,

\begin{eqnarray}
& h_\alpha = h_m + h_\beta & \label{3} \\
& S^z_\alpha = S^z_m + S^z_\beta & \nonumber 
\end{eqnarray}

\noindent where $h_\alpha, h_m, h_\beta$ denote
the number of holes 
and  $S^z_\alpha, S^z_m, S^z_\beta$ denote
the third component of the spin of the corresponding states.

Finally, we can impose the conservation of  the electron-hole
and spin-parity symmetries generated by the operators
$\hat{J}$ and $\hat{P}$, whose action on a $i^{\rm th}$- monomer
is given by \cite{Ramashesa1},

\begin{eqnarray}
& \hat{J}_i | \circ  \rangle = - | \times   \rangle, \;\; 
\hat{P}_i | \circ   \rangle =  - |  \circ  \rangle & \nonumber  \\ 
& \hat{J}_i |  \times   \rangle =   |  \circ  \rangle, \;\;
\hat{P}_i |  \times   \rangle =  | \times    \rangle & \label{4} \\ 
& \hat{J}_i |  \uparrow  \rangle = (-1)^{i+1}  |  \uparrow  \rangle, 
\;\;
\hat{P}_i |\uparrow  \rangle = -  | \downarrow    \rangle & \nonumber \\ 
& \hat{J}_i | \downarrow  \rangle = (-1)^{i+1}   | \downarrow   \rangle, 
\;\; 
\hat{P}_i | \downarrow  \rangle = -   | \uparrow    \rangle & \nonumber  
\end{eqnarray}

The action of  $\hat{J}$ and $\hat{P}$ for a polymer with
$N$ units is simply the tensor product of their actions on
each monomer. 
In the eqs.(\ref{4})  we use the convention according to which 
a state with symmetry  $^1A^+_g$ has $\hat{J}=\hat{P}=1$, 
a state with symmetry $^1B^-_u$ has  $\hat{J}=-\hat{P}= -1$
while a state with symmetry $^3B^+_u$ has  $\hat{J}=-\hat{P}= 1$
( this differs in an overall sign to that used in \cite{Ramashesa1}).
The labels $A$ and $B$ refer to the reflection symmetry 
of the polymer, which  shall not be imposed  explicitely.

Both the monomer states $|m\rangle$ and the block states
$|\alpha \rangle$ transform as follows under 
charge-transfer and spin-parity,

\begin{eqnarray} 
& \hat{J} |\alpha \rangle = \eta^J_\alpha\; |\alpha_J\rangle,
\;\; \hat{J} |m \rangle = \eta^J_m \; |m_J\rangle & \label{5} \\
& \hat{P} |\alpha \rangle = \eta^P_\alpha \; |\alpha_P\rangle,
\;\; \hat{P} |m \rangle = \eta^P_m \; |m_P\rangle & \nonumber 
\end{eqnarray}

\noindent where $\eta^J_m$ and $\eta^P_m$
can be derived from eqs.(\ref{4}),  while  $\eta^J_\alpha$ 
and $\eta^P_\alpha$ are the appropriated ones corresponding
to the type of block chosen. In eq.(\ref{5}) $\alpha_J$ and 
$m_J$ denote the states obtained after the application of
$\hat{J}$ on the states $\alpha$ and $m$ respectively.
All these quantities are  given in tables 1 and 2. 
The MP equation (\ref{1}) preserves the electron-hole 
and spin-parity symmetries provided the MP-amplitudes 
$A_{\alpha,\beta}[m]$ satisfy the following constraints,

\begin{eqnarray}
& A_{\alpha_J,\beta_J}[m_J] = \eta^J_\alpha \eta^J_m \eta^J_\beta
\; A_{\alpha,\beta}[m] & \label{6} \\
& A_{\alpha_P,\beta_P}[m_P] = \eta^P_\alpha \eta^P_m \eta^P_\beta
\; A_{\alpha,\beta}[m] & \nonumber 
\end{eqnarray}

Imposing the spin and charge conservation (\ref{3}),  the 
electron-hole and  the spin-parity symmetries (\ref{6}) 
we are left  with 
a total of 62 non vanishing 
MP-amplitudes $A_{\alpha, \beta}[m]$ out of 576 possible ones.
Moreover only 20 of these 62 parameters are independent.
In table 3 we give a choice for these parameters
in terms of the MP-amplitudes, 
which we shall call hereafter $x_i ( i=1, \dots, 20)$. 
Finally,  the normalization conditions
(\ref{2}) yield three more conditions on the set $x_i$
given by,

\begin{eqnarray}
& x_1^2 + x_2^2 + x_3^2 + 4 x_4^2 + 4 x_5^2 = 1 & \nonumber \\
& x_6^2 + x_7^2 + x_8^2 + 4 x_9^2 + 4 x_{10}^2 = 1 & \label{7} \\
& \sum_{i=11}^{20}    x_i^2 = 1 & \nonumber
\end{eqnarray}

Hence 
altogether
 we are left with 17 independent variational parameters
$y_j (j=1, \dots, 17)$ which will be determined by minimization
of the GS energy. Before we do that it is convenient to parametrized
the $x_i$ parameters in terms of the $y_j$ ones
(see below).
 From physical
reasons we expect that the most important MP-amplitudes 
will be given by
$x_1 = A_{1,1}[{1}], x_8 = A_{2,1}[{3}] $  and 
$x_{17} = A_{3,1}[{9}]$. Indeed,  $x_1, x_8$ and $x_{17}$ 
correspond to the addition
of a singlet, a local $^1B_u^-$ state, and a bonding spin $1/2$ state
 to the GS block $|1 \rangle$ , yielding  a block state with the
same type of symmetry as the monomeric state added. From this
observation the parametrization we are looking for is given by

\begin{eqnarray}
& x_1 = s_1 , x_2 = y_1 s_1, x_3 = y_2 s_1 , x_4 = y_3 s_1 , x_5 = y_4 s_1  &
\nonumber \\
& x_6 = y_5 s_2 , x_7 = y_6 s_2, x_8 = s_2 , x_9 = y_7 s_2 , 
x_{10} = y_8 s_2  & \label{8} \\
& x_{11} = y_9 s_3 , x_{12} = y_{10} s_3, 
x_{13} = y_{11} s_3 , x_{14} = y_{12} s_3 , x_{15} = y_{13} s_3  &
\nonumber \\
& x_{16} = y_{14} s_3 , x_{17} =  s_3, 
x_{18} = y_{15} s_3 , x_{19} = y_{16} s_3 , x_{20} = y_{17} s_3  &
\nonumber \\
& s_1 = 1/\sqrt{ 1 + y_1^2 + y_2^2 + 4 y_3^2 + 4 y_4^2} & \nonumber \\
& s_2 = 1/\sqrt{ 1 + y_5^2 + y_6^2 + 4 y_7^2 + 4 y_8^2} & \nonumber \\
& s_3 = 1/\sqrt{ \sum_{j=9}^{17} y_j^2 } & \nonumber 
\end{eqnarray}

The normalization conditions (\ref{7}) are automatically
satisfied by the parametrization (\ref{8}), which on the
other hand is quite convenient for numerical purposes
\cite{RSDM}. 

If we choose $y_j= 0 ( \forall j)$ then the state
$|1 \rangle_N$ generated by (\ref{1}) consists 
in the coherent superposition of singlets bonds on each
monomer. On the other hand the RR's (\ref{1}) 
also contain 
the Simpson state 
\cite{simpson},
which is the coherent superposition

\begin{equation}
|{\rm Simpson} \rangle_N = \prod_{n=1}^N \;  ( x_1 |1 \rangle_n 
+ x_2 |2 \rangle_n )  
\label{9}
\end{equation}
With this state, the dimerized chain is viewed as a simple one-dimensional
crystal of ethylene where, moreover, the electron correlations are ignored;
this state was the reference state in \cite{RVA-poly}. 
It corresponds to $y_1 = x_2/x_1 \neq 0$
and $y_j = 0 ( {\rm \mbox{for}} \;\; j  > 2)$

\section{Ground state energy}

In this section we shall briefly present the method
for finding the GS energy of the MP ansatz
whose minimization determines the MP parameters
( see references \cite{RSDM} for more details on the 
method). 

Conjugated polymers are customarily  described
by the Pariser-Parr-Pople (PPP) Hamiltonian,
however in our study we shall use a simplified
version of it given by the U-V Hamiltonian
defined as,

\begin{eqnarray}
& H= - t \sum_{i,s} [1 + (-1)^i \delta] (c^\dagger_{i,s} c_{i+1,s}
+ h.c.) & \label{10} \\
& +  U \sum_{i} \; n_{i,\uparrow} n_{i,\downarrow} +
V \sum_i (n_i -1) (n_{i+1} -1) & \nonumber
\end{eqnarray}

\noindent where $c_{i,s}^\dagger$ and $c_{i,s}$ 
are fermionic 
creation and destruction operators at the site
$i$ and spin $s$, $n_{i,s} = c^\dagger_{i,s} c_{i,s}$
and $n_i = n_{i,\uparrow} + n _{i,\downarrow} $. 
We shall work in units where the hopping amplitude
$t$ is set equal  to one. The important parameters
are 
therefore
the dimerization $\delta$, the on-site
Hubbard
coupling $U$ and the nearest neighbour Coulomb interaction
$V$. Since we are working in the monomer basis
it is convenient
to write the  Hamiltonian (\ref{10})
as follows,

\begin{eqnarray}
& H_N = \sum_{j=1}^N h^{(1)}_j + \sum_{j=1}^{N-1} h^{(2)}_{j,j+1}& 
\label{12} \\
& h^{(1)}_j =  - t \sum_{s} [1 + \delta] (c^\dagger_{2j-1,s} c_{2j,s}
+ h.c.) + U (  n_{2j-1,\uparrow} n_{2j-1,\downarrow} +
  n_{2j,\uparrow} n_{2j,\downarrow} ) & \nonumber \\
&   +  
V  (n_{2j -1}-1)  (n_{2j} -1) & \nonumber \\
& h^{(2)}_{j,j-1} =   - t \sum_{s} [1 - \delta] 
(c^\dagger_{2j,s} c_{2j+1,s}
+ h.c.) +  
V  (n_{2j} -1) (n_{2j+1} -1) & \nonumber
\end{eqnarray}

\noindent where $h^{(1)}_j$ is the intramonomer Hamiltonian 
of the $j^{\rm th}$ monomer and $h^{(2)}_{j,j+1}$ is the intermonomer
Hamiltonian coupling the monomers $j$ and $j+1$. 
$N$ denotes the total number of monomers.

The block states $|\alpha\rangle_N$ belong to different
Hilbert
spaces of the Hamiltonian (\ref{12}),  therefore
the vacuum expectation value of $H_N$ will be diagonal
with entries,

\begin{equation}
E^{N}_\alpha = _N\langle \alpha| H_N | \alpha \rangle_N 
\label{13}
\end{equation}

The RR (\ref{1}) yields a RR for these energies given by
\cite{RSDM}

\begin{equation}
E^{(N+1)}_\alpha = \sum_{\beta} \; T_{\alpha, \beta} E^{(N)}_\beta
+ \widehat{h^{(1)}_\alpha} +  \widehat{h^{(2)}_\alpha} \label{14} 
\end{equation}

\noindent where

\begin{eqnarray}
& T_{\alpha, \beta} = \sum_{m} ( A_{\alpha, \beta}[m])^2 & 
\label{15} \\
& \widehat{h^{(1)}_\alpha} = \sum_{m,m',\beta} A_{\alpha,\beta}[m]
A_{\alpha,\beta}[m'] \; \epsilon_1(m,m') & \nonumber \\
& \widehat{h^{(2)}_\alpha} = \sum_{m's,\beta,\beta',\gamma} 
A_{\alpha,\beta}[m_1] A_{\alpha,\beta'}[m'_1] 
A_{\beta,\gamma}[m_2] A_{\beta',\gamma}[m'_2] 
\; \epsilon_2(m_1,m_2;m'_1,m'_2) & \nonumber 
\end{eqnarray}

\begin{eqnarray}
& \epsilon_1(m,m') = \langle m | h^{(1)} | m' \rangle & \label{16} \\
& \epsilon_2(m_1,m_2;m'_1,m_2) = \langle m_2, m_1 | h^{(2)} |
m'_1, m'_2 \rangle & \nonumber  
\end{eqnarray}

The last two expressions are the intramonomer, i.e. $\epsilon_1$, 
and intermonomer, i.e. $\epsilon_2$,  matrix elements
in the monomer basis, which can be computed either analytically
or numerically. 
For the case of the PPP Hamiltonian, the number of these energy
matrix elements (\ref{16}) is huge and it is very lengthy the analytical
computations of so many quantities. Instead, we have used 
{\it numerical} exact diagonalization techniques in order to
compute them numerically once the PPP coupling constants are
specified.
This numerical coding is divided into two parts: 1) We construct
the Hilbert space of states for the one- and two-monomer basis.
This is done in a binary notation using a string of bits of
length $4$ for the one-monomers and $8$ for the two-monomers.
In the first half of each string of bits we encode the spin-up
states and in the second half we encode the spin-downs.
This representation we call it the {\it tensorial basis.}
2) We represent numerically the action of the PPP Hamiltonian 
in the tensorial basis. This facilitates the computation of the
energy matrix elements (\ref{16}).
Lastly, we perform several change of basis to bring the previous
matrix elements to the Valence Bond basis employed in the
variational recurrence relations.

The RR (\ref{14}) can be iterated to give $E^{N}_\alpha$
once $E^1_\alpha$ is known. Actually, the same is true for 
eq.(\ref{1}) which gives the MP states $|\alpha\rangle_N$ 
once $|\alpha \rangle_1$ is given. We shall choose
as initial states   $|\alpha \rangle_1$ the lowest
states of the monomer hamiltonian $h^{(1)}$ in the corresponding
Hilbert space sector. Hence the computation of $E^1_\alpha$
requires the diagonalization of $\epsilon_1(m,m')$.

Now the procedure goes as follows. Using eq.(\ref{14}) we
find the value of $E^N_1$ for a given set of variational
parameters $y_j$ and look for the lowest possible value.
This determines the value of these parameters and correspondingly
that of the MP-amplitudes. One also finds in this way 
the value of the GS energy density per monomer in the thermodynamic
limit,

\begin{equation}
e_\infty = \lim_{N \rightarrow \infty}  E^N_1/N 
\label{17} 
\end{equation}

\section{Results}

In figure 1 we present the GS energies per monomer 
obtained with the MP method outlined above and the
DMRG
for the  cases i) $ U=4,\; V=1, \;  2 \delta= 0.1, 0.3, 0.5, 1.5$ 
and  ii) $ U=3, \; V= 1.2, \;  2 \delta= 0.1, 0.3, 0.5, 1.5$. 
For small dimerizations the relative error of the MP results
as compared with the DMRG is around $4 \%$, while for strong
dimerization it is around $2 \%$.

In figure 2 we plot the absolute value of the 
20 amplitudes $x_i$ described in table 3
for weak dimerization ($\delta = 0.05$) and 
strong dimerization ($\delta = 0.75$) and couplings
$U= 3, V= 1.2$ in both cases. 
It is clear from fig. 2 that for strong dimerization
the MP state is very close to the Simpson state for the
most important amplitudes are $x_1, x_2, x_8$ and $x_{17}$.
For weak dimerization we observe a transfer of weight 
from these parameters to the remaining ones which show
that the charge transfer excitations begin to play 
a more important role. This is specially clear in the behaviour
of $x_4$ which involves the monomer configurations
$(\circ \uparrow + \uparrow \circ), 
(\circ \downarrow + \downarrow \circ), 
(\times \uparrow - \uparrow \times), 
(\times \downarrow - \downarrow \times)$, which are the typical
local CT configurations appearing in the GS. On the contrary
the parameter $x_5$ remains very small showing that
the monomer configurations  
$(\circ \uparrow - \uparrow \circ), 
(\circ \downarrow - \downarrow \circ), 
(\times \uparrow + \uparrow \times), 
(\times \downarrow + \downarrow \times)$ are very unlikely
in the GS.

These results
are encouraging since they show that the MP
approach gives a reasonable representation of the 
GS of the conjugated polymers in terms of a small number
of variational  parameters. They also show the possible
improvements which can be achieved by first rejecting those monomer
configurations which  have small  weight in the GS.
One could also include blocks with spin 1 and singlet blocks
with degeneracy. The latter type of blocks is needed in order
to discuss the interesting  crossing  between the 
energy levels $1 ^1B^-_u$ and $ 2 ^1A^+_g$ \cite{Ramashesa2}.

\section{Conclusions}

This paper represents the first attempt to generate a MP
ansatz of the GS of conjugated polymers. Our results
are rather encouraging since they show that we can get 
new insights and good numerical accuracy by improving 
the ansatz. Unlike other variational methods the MPM
allows for a  systematic improvement,   
becoming  eventually 
exact when keeping a  sufficient number of block states.
Of course in the latter case the method becomes equivalent
to the DMRG one \cite{DMNS}. The 
usefulness
of the MPM thus
lies in a certain compromise between the desired 
numerical accuracy  and the physical insight usually
associated with  the analytic nature   of the method.  
The MPM also demands much less computing effort, an aspect
which is certainly  non negligible

.

{\bf Acknowledgements} We would like to thank useful correspondence with 
S.K. Pati and conversations with S. Ramasesha at
the Max Planck Institute for the Physics of Complex Systems in Dresden 
during the  DMRG98 Seminar/Workshop, at which the present 
work was initiated.

M.A.M.D. and G.S. acknowledge support from the 
DIGICYT under contract No. PB96/0906, 
S.P aknowledges support from the European Commission through the TMR network
contract ERBFNRX-CT96-0079 (QUCEX) and support from the DIGICYT under
contract No. PB96/0906 which permits his stay in Madrid for a short period.

\begin{figure}
\hspace{0.0cm}
\epsfxsize=14cm \epsffile{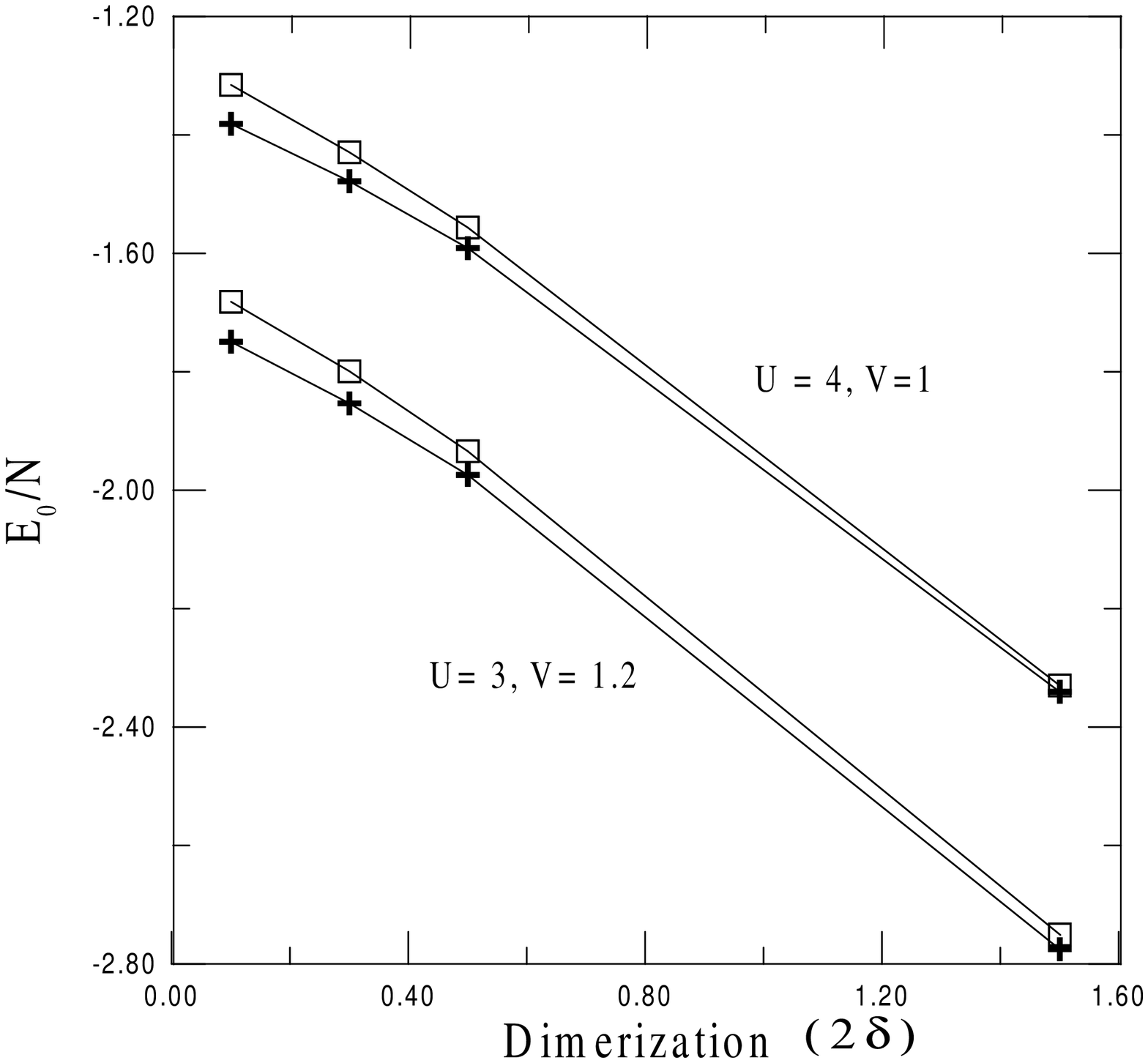}
\narrowtext
\caption[]{ Ground state energies per site using the Matrix Product ansatz
(blank square) compared with DMRG results (solid crosses) plotted against
the degree of dimerization of the polymer chain and for 
several
 values
of the PPP parameters U and V in equation (\ref{10}).
}
\label{fig1} 
\end{figure}
\noindent

\begin{figure}
\hspace{0.0cm}
\epsfxsize=14cm \epsffile{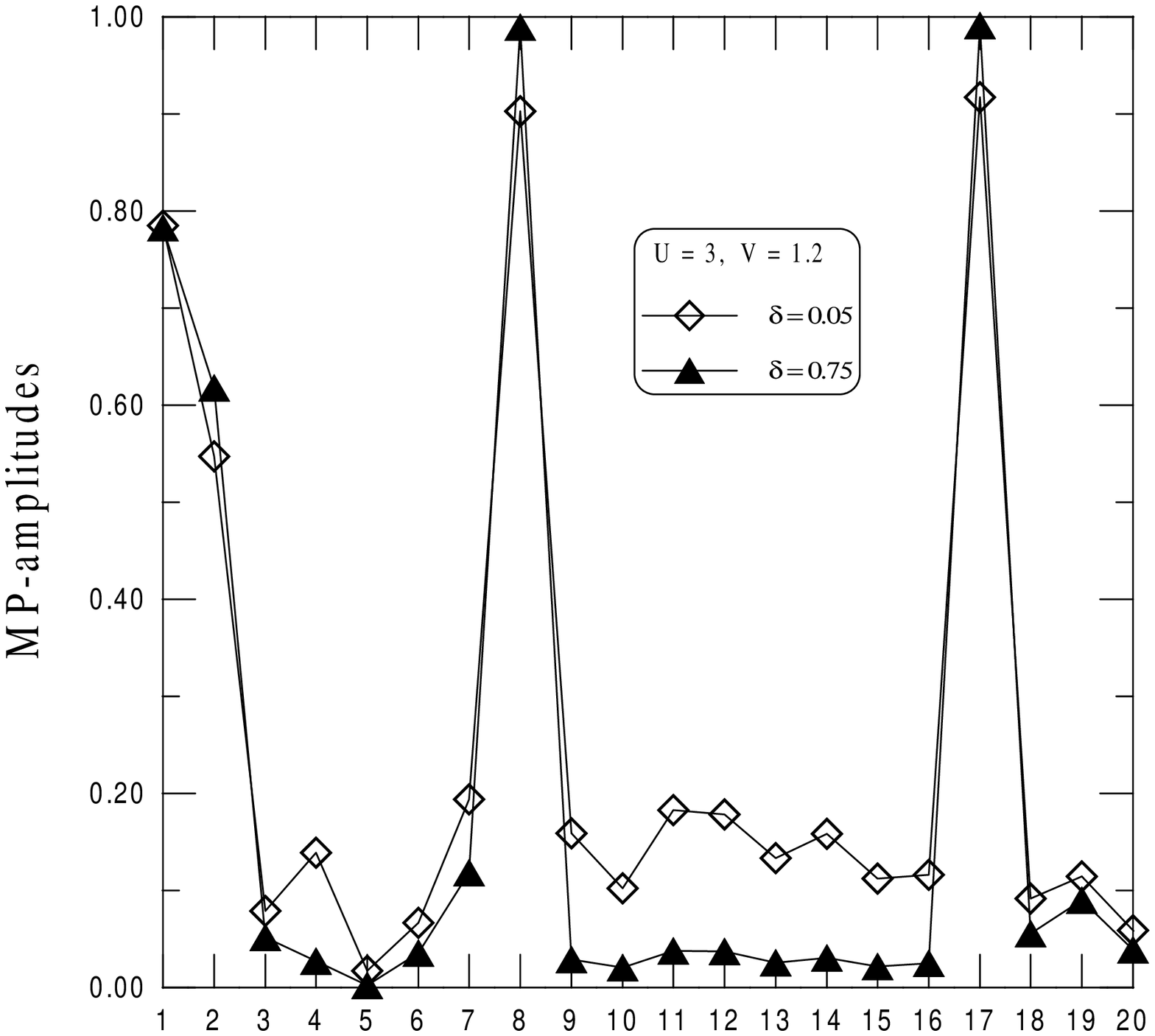}
\narrowtext
\caption[]{ Absolute value of the twenty independent
variational parameters used 
in the Matrix Product
ansatz using six block bulk states.}
\label{fig2} 
\end{figure}
\noindent

\begin{center}
\begin{tabular}{|c|c|r|r|r|r|r|r|} 
\hline
\hline
State  & $m$  & $h$ & $2 S^z$ & ${m}_J$ & 
$\eta^J_m$ & $m_P$ & $\eta^P_m$   \\ 
\hline
\hline
$\bullet - \bullet $ & 1 & 0 & 0 & 1 & 1 & 1 & 1 \\
\hline
$\times \circ + \circ \times$ & 2 & 0 & 0 & 2 & 1 & 2 & 1 \\
\hline 
$\times \circ - \circ \times$ & 3 & 0 & 0 & 3 & -1 & 3 & 1 \\
\hline 
$\uparrow \uparrow$ & 4 & 0 & 2 & 4 & 1 & 6 & -1 \\
\hline
$\uparrow \downarrow + \downarrow \uparrow $ & 5 & 0 & 0 & 5& 1& 5 & -1 \\
\hline
$\downarrow \downarrow $ & 6 & 0 & -2 & 6 & 1 & 4 & -1 \\
\hline
$\circ \circ $ & 7 & 2 & 0 & 8 & -1 & 7 & -1 \\
\hline
$\times \times $ & 8 & -2 & 0 & 7 & -1 & 8 & -1 \\
\hline 
$\circ \uparrow + \uparrow \circ$ & 9 & 1 & 1 & 15 & -1 & 10 & -1 \\
\hline
$\circ \downarrow + \downarrow \circ$ & 10 & 1 &- 1 & 16 & -1 & 9 & -1 \\
\hline
$\circ \uparrow - \uparrow \circ$ & 11 & 1 & 1 & 13 & -1 & 12 & -1 \\
\hline
$\circ \downarrow - \downarrow \circ$ & 12 & 1 &- 1 & 14 & -1 & 11 & -1 \\
\hline
$\times \uparrow + \uparrow \times$ & 13 & -1 & 1 & 11 & 1 & 14 & 1 \\
\hline
$\times \downarrow + \downarrow \times$ & 14 & -1 &- 1 & 12 & 1& 13 &  1 \\
\hline
$\times \uparrow - \uparrow \times$ & 15 & -1 & 1 & 9 & 1 & 16 & 1 \\
\hline
$\times \downarrow - \downarrow \times$ & 16 & -1 &- 1 & 10 & 1& 15 &  1 \\
\hline
\hline
\end{tabular}
\end{center}
\begin{center}
Table 1.- States forming the monomer basis of the MP ansatz. 
The states given in the first column
have to be  normalized. $\bullet - \bullet$ represents
a singlet valence bond state, $\times $ represents a double
occupied site, $\circ$ symbolizes an empty site and 
$\uparrow, \downarrow$ 
symbolizes
singly occupied sites with spin up and down.
$h$ denotes the excess or defect of  holes   
as compared to the half filling situation. $2 S^z$ is twice
the third component of the spin. $m_J$ and $m_P$ are the states
obtained upon applying the operators $\hat{J}$ and $\hat{P}$ 
on the monomer state $m$ defined in eqs. (\ref{4}). 
$\eta^J_m$ and $\eta^P_m$ are the corresponding  signs 
appearing in eqs.(\ref{5}). 
\end{center}

\begin{center}
\begin{tabular}{|c|c|r|r|r|r|r|r|} 
\hline
\hline
State  & $\alpha$  & $h$ & $2 S^z$ & ${\alpha}_J$ & 
$\eta^J_\alpha$ & $\alpha_P$ & $\eta^P_\alpha$   \\ 
\hline
\hline
$\bullet - \bullet $ & 1 & 0 & 0 & 1 & 1 & 1 & 1 \\
\hline
$\times \circ - \circ \times$ & 2 & 0 & 0 & 2 & -1 & 2 & 1 \\
\hline 
$\circ \uparrow + \uparrow \circ$ & 3 & 1 & 1 & 5 & -1 & 4 & -1 \\
\hline
$\circ \downarrow + \downarrow \circ$ & 4 & 1 &- 1 & 6 & -1 & 3 & -1 \\
\hline
$\times \uparrow - \uparrow \times$ & 5 & -1 & 1 & 3 & 1 & 6 & 1 \\
\hline
$\times \downarrow - \downarrow \times$ & 6 & -1 &- 1 & 4 & 1& 5 &  1 \\
\hline
\hline 
\end{tabular}
\end{center}
\begin{center}
Table 2. The notations are as in table 1. The states appearing in the first
column are for illustration purposes. They simply show the type of symmetry
of the block state as compared with the monomer states defined in table 1. 
\end{center}

\begin{center}
\begin{tabular}{|r|r|r|r||r|r|r|r|} 
\hline
\hline
$x_i$   & $\alpha$  & $m$ & $\beta$ & $x_i$ & $\alpha$  & $m$ & $\beta$  \\
\hline
\hline
$x_1$  &  1 & 1 & 1 & $x_{11}$ & 3  & 1 & 3   \\
\hline
$x_{2}$   & 1   & 2  & 1  & $x_{12}$  & 3  &2  &3      \\
\hline  
$x_3$   & 1   & 3  & 2 & $x_{13}$  & 3  &3  &3      \\
\hline  
$x_4$   & 1  & 9  & 6 & $x_{14}$  & 3  & 4 & 4     \\
\hline  
$x_5$   & 1  & 11  & 6 & $x_{15}$  & 3  &5  &3      \\
\hline  
$x_6$   & 2  & 1  & 2 & $x_{16}$  &3  &7  & 5     \\
\hline  
$x_7$   & 2   & 2  &2  &$x_{17}$ & 3  &9  &1      \\
\hline  
$x_8$   & 2  & 3  & 1 & $x_{18}$ &3  & 9 & 2     \\
\hline  
$x_9$   & 2  & 9  & 6 & $x_{19}$ &3  &11  & 1     \\
\hline  
$x_{10}$   & 2  & 11  & 6  &$x_{20}$  &3  & 11 & 2      \\
\hline  
\hline 
\end{tabular}
\end{center}
\begin{center}
Table 3. List of the variational parameters $x_i$ in terms of 
the MP-amplitudes $A_{\alpha,\beta}[m]$. The total of non vanishing
amplitudes $A_{\alpha,\beta}[m]$ is 62. The remaining 42 amplitudes 
can be computed  using eqs.(\ref{6}). 
\end{center}




\begin{thebibliography}{99}

\bibitem{review}
D. Baeriswyl, D.K. Campbell and S. Mazumdar, in Conjugated Conducting
Polymers, edited by H. Kiess (Springer-Verlag, Heidelberg, 1992), pp 7-133.

\bibitem{huckel}
E. H\"{u}ckel, Z. Physik {\bf 76}, 628 (1932).

\bibitem{ssh}
W.P. Su, J.R. Schrieffer and A.J. Heeger, Phys. Rev {\bf B22}, 2099
(1980).

\bibitem{holstein}
T. Holstein, Ann. of Phys. {\bf 8}, 325 (1959).

\bibitem{ppp1}
R.\ Pariser and R.\ G.\ Parr, J.\ Chem.\ Phys.\ 21 (1953) 446; J.\ A.\ Pople, Trans.\ 
Faraday Soc.\ 42 (1953) 1375.

\bibitem{ppp2} R.G. Parr, {\it The Quantum Teory of Molecular Electronic
Structure}, Benjamin, New York, 1958.


\bibitem{white} S.R. White, Phys. Rev. Lett. 69, 2863 (1992),
Phys. Rev. B 48, 10345 (1993).



\bibitem{Ramashesa1} S. Ramashesa, S. K. Pati, 
H. R. Krishnamurthy, Z. Shuai and J.L. Bredas,
Phys. Rev. B 54, 7598 (1996). 


\bibitem{ramasesha} 
Z. Shuai, S.K. Pati, W.P. Su, J.L. Bredas and S. Ramasesha 
Phys. Rev. B {\bf 55} 15368 (1997).


\bibitem{bb}
W. Barford and R.J. Bursill, Chem. Phys. Lett.{\bf 268} 535 (1997);
Synth. Met. 85 1155 (1997).
R. J. Bursill, C. Castleton and W. Barford, Chem. Phys. Lett.
{\bf 294} 305 (1998).
M. Boman and R.J. Bursinll, Phys. Rev. {\bf B 57} 15167 (1998).

\bibitem{yaron}
D. Yaron, E.E. Moore, Z. Shuai, J.J. Bredas, 
J. Chem. Phys. {\bf 108}, 7451 (1998).


\bibitem{fano} G. Fano, F. Ortolani and L. Ziosi,
J. Chem. Phys {\bf 108}, 9246 (1998);
J. Chem. Phys {\bf 110}, 1277 (1999).




\bibitem{pleutin}
S. Pleutin and J.L. Fave, J. Phys. Cond. Matt. {\bf 10}, 3941 (1998).


\bibitem{RVA-poly}
S. Pleutin, E. Jeckelman, M.A. Martin-Delgado and G. Sierra,
preprint July 1999, cond-mat/9908062, 
to appear in 
Prog. Theo. Chem. Phys.



\bibitem{RVA} G. Sierra and M.A. Martin-Delgado, 
Phys. Rev. B 56, 8774 (1997). For a review on
the RVA method see M.A. Martin-Delgado and G. Sierra
in  ``Density Matrix Renormalization Group'', eds. I. Peschel
et al. LNP vol. 528. Springer-Verlag, 1999.









\bibitem{exciton-basis} M. Chandros, Y. Shimoi and S. Mazumdar,
Phys. Rev. B 59, 4822 (1999). 


\bibitem{matrix-product} A. Klumper, A. Schadschneider
and Z. Zittartz, Europhys. Lett. 24, 293 (1993);
S. Ostlund and S. Rommer, Phys. Rev. Lett. 75, 3537 (1995);
 S. Rommer and S. Ostlund, Phys. Rev. B 55, 2164 (1997).


\bibitem{RSDM} J.M. Roman, G. Sierra, J. Dukelsky and M.A.
Martin-Delgado, J. Phys. A: Math. Gen. 31, 9729 (1998).

\bibitem{simpson} W.T. Simpson, J. Am. Chem. Soc. {\bf 77}, 6164 (1955).

\bibitem{Ramashesa2} Z. Shuai, J.L. Bredas,
S.K. Pati and S. Ramashesa, Phys. Rev. B 56, 9298 (1997). 



\bibitem{DMNS} J. Dukelsky, M.A. Martin-Delgado, T. Nishino 
and G. Sierra, Europhys. Lett. 43, 457 (1998).

\end{thebibliography}
\end{document}